\documentclass[11pt]{article}
\usepackage{graphicx}

\columnwidth\textwidth

\def\fii{\varphi}
\def\al{\alpha}

\def\ro{\varrho}

\def\d{\partial}
\def\=d{\,{\buildrel\rm def\over =}\,}

\def\sqr#1#2{{\vcenter{\vbox{\hrule height.#2pt\hbox{\vrule width.
#2pt height#1pt \kern#1pt \vrule width.#2pt}\hrule height.#2pt}}}}
 
\def\eps{\varepsilon}

\def\te{\vartheta}
\def\B{\Bigl}

\begin{document}

\title{DIPOLE DENSITY INSTEAD OF POTENTIALS IN ELECTROCARDIOLOGY }

\author{G\"unter Scharf\footnote{e-mail: scharf@physik.uzh.ch}
\\ Physics Institute, University of Z\"urich
\\Lam Dang\footnote{e-mail: lam.dang.ch@gmail.com}
\\HerzGef\"assZentrum, Klinik im Park
\\8022 Z\"urich}

\date{ }

\maketitle\vskip 3cm

\begin{abstract} 

We discuss the forward and inverse problems between the potential $V(x)$ measured in a heart chamber and its sources represented by a dipole density $d(y)$ located on the heart wall. We show that the mapping from $d(y)$ to $V(x)$ is a compact integral operator. Its inverse is unbounded which makes the inverse problem ill-posed in the mathematical sense. We investigate methods to solve the inverse problem approximately in view of the mapping of complicated cardiac arrhythmias. We point out an analogy between phase mapping and 2-dimensional hydrodynamics.

\end{abstract}

\newpage
\section{Introduction}

Electrocardiology rests upon the study of electric fields generated by the heart. As a physicist one immediately asks: What are the sources of these fields. This question has two answers. (i) On the microscopic scale the sources are $K^+$ and $Na^+$ ions and negatively $Cl^-$ ions and proteins. But there is no large separation of positive and negative charges. The membrane of the active cells is able to open channels for the positive ions only, the negative ones remain confined in the cell, the result is a microscopic dipole. (ii) On the macroscopic scale, where individual cells cannot be resolved, we then have a macroscopic dipole density but no charge density (monopole). The variation of this dipole density in space and time spreads through the tissue as a propagating wave of depolarization.

If the dipole density as the source is known it is straightforward to calculate the corresponding electric potential which can be measured. This is the so-called forward problem which in mathematical terms is the solution of Poisson's equation. However, the desired medical information is given by the sources, i.e. by the macroscopic dipole density. Therefore we must solve the inverse problem of calculating the dipole density from potential measurements. In the past most authors have considered a different inverse problem, namely the determination of the potential near the heart wall from potential measurements in the heart chamber or on the body surface. In this case no assumption about the sources is made, one solves a boundary-value problem for Laplace's equation. However, the potential near or on the heart wall is a superposition of the fields generated by near and distant active regions. As a consequence the potential only presents a broad and smooth depiction of the local electric activity.

From these arguments it is clear that the dipole density is better suited for the mapping of complicated arrhythmias than the potential. One may ask why this approach has not been tried before, to be correct: almost not. There is a pioneering work by A. van Oosterom [1] who always tried to model the sources instead of calculating potentials. He already discusses the scalar dipole density as a model of cardiac activity under the name ``equivalent double layer model'' (EDL). The reason why other authors did not follow this route may be the fact that the dipole density is harder to obtain than the contact potential at the heart wall, we shall discuss this in the next section. To compute the dipole density requires much better input data. Indeed, four years ago we have made test calculation with data which were collected by the EnSite system of St.Jude. The results were unsatisfactory. Only the new device from Acutus Medical Inc. called  AcQMap System seems to be suited for our purpose. It has a recording catheter with 48 electrodes plus 48 ultrasound transducers for distance measurements. It allows determination  of the heart chamber geometry and the potential measurements simultaneously at the same time. This eliminates the error due to the motion of the measuring electrodes and of the heart itself.

The paper is organized as follows. In the next section we discuss the forward and inverse problems for the dipole density. We show that the mapping between the dipole density and the potential is a compact integral operator which can well be represented by a finite dimensional matrix. But its inverse is an unbounded operator. This has the bad consequence that the inverse problem is a so-called ill-defined problem which requires special techniques of solution. These facts are illustrated in section 3 by a simple solvable model which we also use to test the general numerical code for solving the inverse problem for the dipole density. In the last section we consider phase mapping which seems to be the suitable technique to map complicated arrhythmias as atrial fibrillation. We compare the dynamics of phase singularities with the vortex dynamics in 2-dimensional hydrodynamics.

\section{Forward and inverse problem for the dipole density}

We apply macroscopic electrodynamics in matter to the heart filled with blood. At this point one usually refers to Jackson [2] and starts from the phenomenological Maxwell's equations
$${1\over c}{\d D\over\d t}={\rm curl}\,H-{4\pi\over c}j,\quad {\rm div}\,D=\ro\eqno(2.1)$$
$${1\over c}{\d B\over\d t}=-{\rm curl}E,\quad {\rm div}B=0.\eqno(2.2)$$
Unfortunately Jackson does not make a clear distinction between microscopic electrodynamics in vacuum and macroscopic electrodynamics in matter. So we prefer the concise book by the first author [3] and refer to the derivation of the macroscopic Maxwell's equations given there. Since the temporal variation of the cardiac fields is slow compared to the propagation of the fields through the body (with the speed of light), we can use the quasi-static approximation and neglect the time derivatives of $D$ and $B$.
Then by (2.1) the conduction current $j(x)$ is divergence-less
$${\rm div}\,j(x)=0\eqno(2.3)$$
and by (2.2) the electric field is curl-less
$${\rm curl}\,E(x)=0.\eqno(2.4)$$
This implies that the electric field has a scalar potential
$$E={\rm grad}\,V.\eqno(2.5)$$
The blood in the heart is a homogeneous medium with constant conductivity $\sigma$, so that $j=\sigma E$. Substituting this into equation (2.3) yields
${\rm div}E=0$. Then equation (2.5) implies Laplace's equation
$${\rm div\, grad}\,V=\triangle V=0.\eqno(2.6)$$
In the quasi-static approximation the time dependence of the fields has completely disappeared. One calculates the potential or dipole density at fixed time, and then makes a movie for successive time instants.
Summing up we describe the heart as a uniform volume conductor [4] with electric dipole sources in the heart wall.

The Laplace equation holds inside the heart chamber filled with blood. In the whole body the situation is much more complicated. Here we have large inhomogeneities (lunges and bones) so that $\sigma$ is no longer constant. If one measures potentials on the body surface (ECG) one must construct a detailed model of the body, in order to derive the sources from those data. To avoid this severe problem we assume that we have a multi-electrode catheter in the heart chamber which measures the potential $V(t,x)$ at various locations. In the heart wall we have the dipole sources which we assume to be localized on a 2-dimensional
surface $S$. Then instead of (2.6) we have Poisson's equation of the form
$$\triangle V(x)=-4\pi{\rm div}(nd(x)\delta_S(x))\eqno(2.7)$$
where $\delta_S(x)$ is the Dirac measure on $S$, $n$ is the outer normal at point x on the surface and $d(x)$ is the surface dipole density
(dipole strength per area). Such a source is also called a dipole layer or double layer in electrostatics. The direction of the dipole moment is
normal to the surface S. Note that this source describes the $macroscopic$ dipole density. The microscopic dipoles can have different directions, but
this microscopic structure can hardly be resolved in detail by macroscopic non-contact techniques. The solution of (2.7) for $V$ is given by the surface integral
$$V(x)=\int\limits_S d(y){\d\over\d n_y}{1\over |x-y|}dS_y\eqno(2.8)$$
where $\d/\d n_y$ is the derivative in the normal direction at point $y$ on $S$ and $dS_y$ is the surface measure. This integral can be rewritten as follows
$$V(x)=\int\limits_S d(y){\cos\fii_{xy}\over |x-y|^2}dS_y\eqno(2.9)$$
where $\fii_{xy}$ is the angle between the vector $x-y$ and the normal $n$ [5]. If the dipole density $d(y)$ is 
given the calculation of the potential $V(x)$ is straightforward, this is the forward problem.

The inverse problem of computing $d(y)$ from measured $V(x)$ is much harder. The reason is the following. The integral operator (2.8) which maps $d(y)$
to the potential $V(x)$ is a $compact$ operator in the mathematical sense. This important fact must be proved.

{\bf Proof:}

 Here we follow the best mathematical reference we know [6]. Let the heart wall be a closed smooth surface $S$ where the dipole density $d(y)$ is located. This is no serious restriction because at the valves $d(y)$ can be put equal to 0. Let $S'$ be another smooth closed surface completely inside the blood volume where electrodes are placed to measure the potential $V(x)$. The dipole density $d(y)$ is assumed to be bounded and continuous on $S$. This implies that $V(x)$ is bounded and continuous on $S'$, because the kernel in (2.9) is continuous (note that $y\in S$ and $x\in S'$ so that we have always $x\ne y$). As usually let $C(S)$ be the Banach space of bounded continuous functions on $S$ and similarly $C(S')$. Then the integral operator (2.9) maps $C(S)$ on $C(S')$ and to simplify the notation we write it as
$$V(x)=\int\limits_S d(y)\,K(x,y)\,dS_y.\eqno(2.10)$$
This is a bounded operator $K$ with the operator norm
$$\Vert K\Vert=\max_{x\in S'}\int\limits_S\vert K(x,y)\vert dS_y.\eqno(2.11)$$

To prove that $K$ is even compact we must use a decomposition of unity. This is a sequence of positive continuous functions $e_j(x)$ with compact support on $S'$ with
$$\sum_{j=1}^ne_j(x)=1$$
and the following property: for every compact set $M\subset S'$ the intersection of $M$ and the support of $e_j$ is not empty for finitely many $j$, only. Since the kernel $K(x,y)$ is uniformly continuous on $S\times S'$ there exists a decomposition of unity and points $x_j$ in the support of $e_j$ such that
$$\vert K(x,y)-\sum_{j=1}^n e_j(x)K(x_j,y)\vert <\eps\eqno(2.12)$$
for all $x\in S'$, $y\in S$ and arbitrary $\eps$. The approximating integral operator $K_\eps$ defined by the sum of product kernels in here is clearly compact. The approximation is in the operator norm because
$$\Vert K-K_\eps\Vert=\max_x\int\limits_{S'}\vert K(x,y)-\sum_{j=1}^n e_j(x)K(x_j,y)\vert dS_y <\eps\vert S'\vert\eqno(2.13)$$
where $\vert S'\vert$ is the total area of $S'$. This proves that $K$ is the limit of a converging sequence of compact operators and, therefore, compact.

{\bf End of proof.}.

This fact has one good
and one bad consequence. The good one is that compact operators can well be approximated by finite-dimensional matrices. Then (2.9) becomes a matrix
equation
$$V(x_j)=\sum_kW_{jk}d_k.\eqno(2.14)$$
The solution of the inverse problem is then given by the inverse matrix
$$d_k=\sum_jW_{kj}^{-1}V(x_j).\eqno(2.15)$$
Compact operators have infinitely many eigenvalues $\lambda_n$ which accumulate only at 0, $\lambda_n\to 0$ for $n\to\infty$ (theorem of F.Riesz [6]). As a consequence the approximating matrix $W$ in (2.14) has small eigenvalues and this is unavoidable. In the inverse (2.15) we then have $\lambda_n^{-1}\to\infty$, so that the corresponding part in the data $V(x_j)$ gets strongly amplified. This is the ill-posed nature of the inverse problem. To avoid a huge amplification
of the noise one must cut off the smallest eigenvalues. This is a convenient regularization method called truncated singular value decomposition (TSVD).

Another widely used method of regularization is the one of Tikhonov [7]. To solve the linear equation (2.14) $Wd - V = 0$ one considers the variation principle
$$(Wd-V,\,WKd-V)+\gamma (Rd,\,Rd)=\Phi(d)=\min,\eqno(2.16)$$
where $R$ is a ``regularizing ''operator (mostly $R=1$) and $\gamma$ is the regularization parameter. Putting the variational derivative $\Phi'(d)$ equal to 0 we obtain the equation
$$(W^+W+\gamma R^+R)d=W^+V\eqno(2.17)$$
where the cross means the adjoint operator. For positive $\gamma$ the accumulation of eigenvalues at 0 is removed in the operator on the left. The latter can be inverted (the inverse is bounded) and the dipole density can be computed. The advantage of this regularization method is that the regularization parameter $\gamma$ can be varied continuously. The optimal choice of $\gamma$ is a serious problem which is discussed in the next section. For $R=1$ one has the first order Tikhonov regularization. 

The inverse problem of electrocardiology in the standard sense is a voltage to voltage approach where one calculates the potential $V(y_S)$ at points $y_S$ on or near the wall $S$. If the dipole density $d$ is known this is a forward calculation (2.9)
 $$V(y_S)=Ld\eqno(2.18)$$
 with a new integral operator $L$ because $y_S$ is now on or near the surface. If $y_S$ is on the wall $S$ the kernel of $L$ has a singularity at $y=y_S$. Nevertheless $L$ is still compact because it is again the limit of compact operators (corresponding to a sequence of surfaces $S'$ converging to $S$ from the interior). Using $d=K^{-1}V$ with the unbounded operator $K^{-1}$, we can eliminate $d$ and get 
$$V(y_S)=LK^{-1}V(x).\eqno(2.19)$$
This operator is better behaved than $K^{-1}$ alone, because $L$ damps the large eigenvalues of $K^{-1}$. Therefore the standard voltage inverse problem is easier to solve, which means that less strong regularization is necessary. But as discussed in the introduction, it gives less precise  information on the electrical activity of the heart.

\section{A solvable model and the numerical code}

To compare dipole density and potential and test candidate regularization methods, a mathematical model with absolutely known dipole density $d(y)$ was defined for which the corresponding potential $V(x)$ can be calculated exactly. This model represents dipole density ``frozen'' at one instant of time. A simple solvable model is obtained as follows.
Let a sphere of radius 1 represent the heart wall (endocardial surface) $S$ and choose the dipole density applied upon it according to the formula $$d(y)=d_0\exp(-p\cos\te)\eqno(3.1)$$
where $\te$ is the polar angle with respect to the $z$-axis, $p$ is a positive parameter and $d_0$ a normalization factor. This distribution is rotationally symmetric around the $z$-axis, it has a maximum at the south pole $\te=\pi$ and diminishes toward the north pole.
A large value of the parameter $p$ causes the maximum-region on the south pole to be narrow, whereas a small value of $p$ causes the maximum-region to be broad. The normalization factor $d_0$ conveniently scales the density values so that the integral over the unit sphere is equal to 1:
$$d_0={p\over \exp{p}-\exp{(-p)}}.\eqno(3.2)$$

The corresponding voltage $V(x)$ for this dipole density can be exactly calculated as follows. We expand $d(y)$ in terms of Legendre polynomials $P_l(\cos\te)$  with respect to the polar angle $\te$ using the integral
$$\int\limits_{-1}^{1}e^{-p\xi}P_l(\xi)d\xi=(-1)^l\sqrt{{2\pi\over p}}I_{l+1/2}(p)\eqno(3.3)$$
where $I_{l+1/2}$ is the modified spherical Bessel function. Then we obtain
$$d(y)=\sum_{l=0}^\infty y^lD_lP_l(\cos\te)\eqno(3.4)$$
with
$$D_l=(-1)^l(2l+1)\sqrt{{\pi p\over 2}}{e^p\over e^{2p}-1}I_{l+1/2}(p).\eqno(3.5)$$

On the other hand the kernel in the potential integral (2.8) can also be expanded in terms of Legendre polynomials. We start from the well-known expansion
$${1\over \vert x-y\vert}={1\over y}\sum_{l=0}^\infty\B({r\over y}\B)^lP_l(\cos\al)\eqno(3.6)$$
where $r=\vert x\vert$, $\al$ is the angle between the vectors $x$ and $y$ and $\vert x\vert<\vert y\vert$. The normal derivative $d/dn$ in (2.8) on the unit sphere is equal to $d/dy$, hence
$${d\over dn_y}{1\over\vert x-y\vert}=-\sum_l(l+1){r^l\over y^{l+2}}P_l(\cos\al).\eqno(3.7)$$
Substituting this into (2.8) we arrive at
$$V(x)=-\sum_l (l+1)r^l\int\limits_S d(y){P_l(\cos\al)\over y^{l+2}}dS_y.\eqno(3.8)$$
A general bounded continuous dipole density on the unit sphere can be expanded in terms of spherical harmonics
$$d(y)=\sum_ly^l\sum_{m=-l}^lD_{lm}Y^m_l(\te,\fii).\eqno(3.9)$$
Inserting this into (3.8) and using the following integral over the angles
$$\int Y_{l'}^{m'}(\te',\fii')P_l(\cos\al)d\cos\te' d\fii'={4\pi\over 2l+1}\delta_{ll'}Y_l^{m'}(\te,\fii)\eqno(3.10)$$
we get the desired potential in the form
$$V(x)=-4\pi\sum_l{l+1\over 2l+1}r^l\sum_mD_{lm}Y_l^m(\te,\fii).\eqno(3.11)$$
This general result can be applied to our rotationally symmetric dipole density (3.4) which gives the potential everywhere in the unit sphere:
$$V(x)=\sum_l x^lV_lP_l(\cos\te)\eqno(3.12)$$
with
$$V_l=-4\pi (-1)^l(l+1)I_{l+1/2}(p)\sqrt{{\pi p\over 2}}{e^p\over e^{2p}-1}.\eqno(3.13)$$
The sum over $l$ is rapidly converging so that one can stop at a finite value $l_{\rm max}$ and gets any desired accuracy.

In Figure A we compare $d(x_S)$ and $V(x_S)$ for $p=5$. We have  normalized both quantities to a maximum value 1 for the purpose of direct comparison. As we follow both distributions from maximum (south pole) toward the minimum (north pole), the voltage has a long rightward tail (power law), compared to the rapid (exponential) descent of the dipole density. This shows clearly the local nature of the dipole density in contrast to the broad distribution of the potential.

\begin{figure}
\centering
\includegraphics[width=\textwidth]{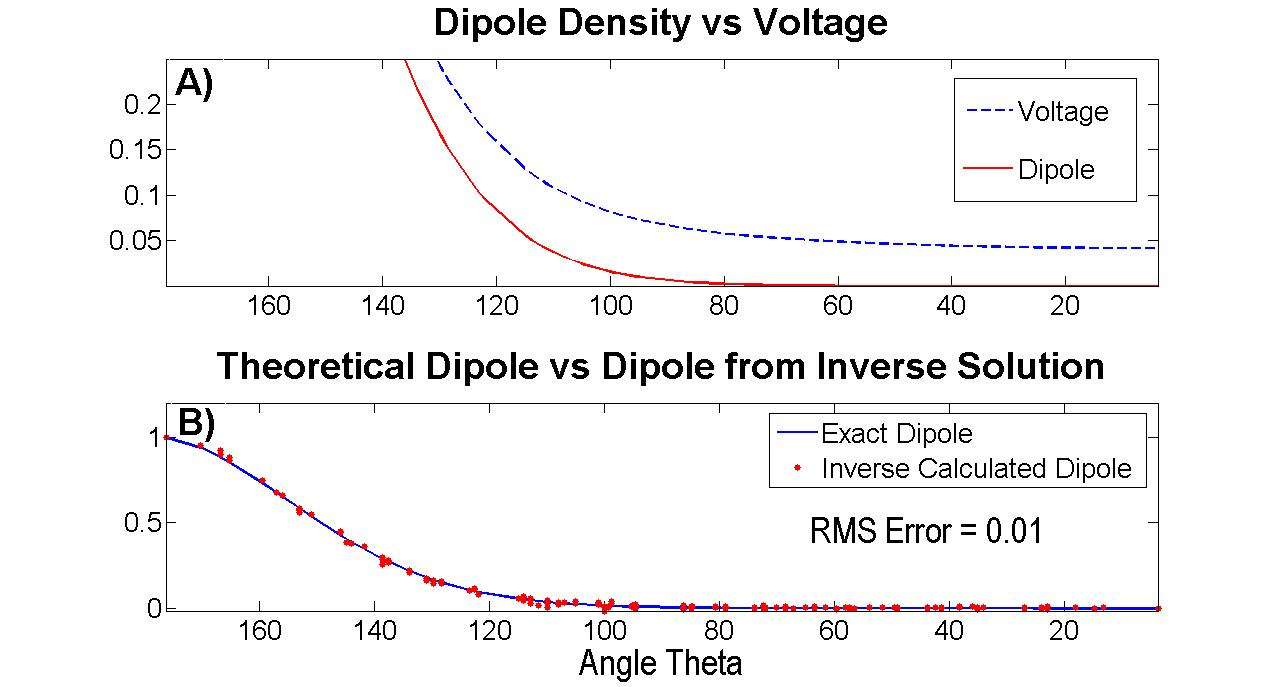}
\caption{ Dipole density and voltage on the wall as function of the polar angle.}
\end{figure}

Next we consider the general numerical code. The heart wall $S$ is covered by a triangular mesh. Since we want to have a continuous dipole density $d(y)$ we approximate it by piecewise linear functions
$$h_n(y)={{\rm det}(x_k, x_l,y)\over {\rm det}(x_k,x_l,x_n)},\quad y\in\triangle_{kln}.\eqno(3.14)$$
and zero otherwise. Here $x_k$, $x_l$, $x_n$ are the vectors of the corner points of the triangle $\triangle_{kln}$ and det is the $3\times 3$ determinant of the 3 vectors. These functions satisfy the condition
$$h_n(x_m)=\delta_{mn}.\eqno(3.15)$$
The unknown dipole density is expanded in the form
$$d(y)=\sum_{n=1}^N d_nh_n(y).\eqno(3.16)$$
Then the integral equation (2.8) becomes the following set of linear equations for $N$ unknowns $d_n$
$$V(x_m)=\sum_{n=1}^N W_{mn}d_n,\quad m=1,\ldots N\eqno(3.17)$$
where
$$W_{mn}=-\int\limits_S h_n(y)\nabla_y{1\over\vert y-x_m\vert}\cdot dS_y.\eqno(3.18)$$

The integral (3.18) is a sum of integrals over the triangles with corner $n$. These integrals can be calculated analytically [8].
The result for one triangle $\triangle_{kln}$ is equal to
$$M_{mn}={1\over A^2}\B[\vec z_n\cdot\vec n\Omega+d(\vec y_k-\vec y_l)\cdot\vec S\B].\eqno(3.19)$$
Here
$$\vec y_k=x_k-x_m,\quad \vec y_l=x_l-x_m,\quad y_n=x_n-x_m$$
$$\vec z_n=\vec y_k\times \vec y_l,\quad d=\vec y_k\cdot(\vec y_l\times\vec y_n),\eqno(3.20)$$
and $\vec n$ is the normal of the triangle and $A$ its absolute value
$$\vec n=(\vec y_l-\vec y_k)\times (\vec y_n-\vec y_k),\quad A=\vert\vec n\vert=2F\eqno(3.21)$$
where $F$ is the area of the triangle. The vector $\vec S$ is given by
$$\vec S=(\vec y_k-\vec y_l)\gamma_k+(\vec y_l-\vec y_n)\gamma_l+(\vec y_n-\vec y_k)\gamma_n\eqno(3.22)$$
with
$$\gamma_k={1\over\vert\vec y_k-\vec y_l\vert}\log{{\vert\vec y_l\vert\vert\vec y_l-\vec y_k\vert+\vec y_l\cdot (\vec y_l-\vec y_k)
\over \vert\vec y_k\vert\vert\vec y_l-\vec y_k\vert+\vec y_k\cdot (\vec y_l-\vec y_k)}}\eqno(3.23)$$
and cyclic $k,l,n$. Finally $\Omega$ is the solid angle of the triangle subtended at the view point $x_m$. A convenient formula for $\Omega$ has been given by van Oosterom and Strackee [9].

If we substitute $h_n$ in (3.18) by 1, we obtain the so-called Gauss-integral which is equal to the solid angle $\Omega$. Since $S$ is a closed surface and $x_m$ is in the interior we get $4\pi$. This leads to the sum rule
$$\sum_n W_{mn}=4\pi\eqno(3.24)$$
which holds exactly because the discrete triangulated surface subtends the same solid angle $4\pi$. The sum rule is an important test of the code,
it must be satisfied with machine accuracy. In other words, the matrix $W_{mn}$ is a stochastic matrix times $4\pi$, it has an eigenvector $d=(1,1,\ldots)$ with eigenvalue $4\pi$. Unfortunately, it also has very small eigenvalues because it approximates a compact operator. Then the inverse problem requires regularization. This leads to some error in the resulting dipole density. We now discuss this essential problem in our solvable model. 

To be near reality we define a spherical ``basket'' of radius r=0.5 which we first place concentric with the unit sphere representing the heart wall. We calculate the exact potential values at 186 evenly distributed points on this basket. This represents the measured values on an array of electrodes of a basket catheter. Finally, we calculate the dipole density on the heart wall ($r=1$) by solving the inverse problem
$$d=W^{-1}V\eqno(3.25)$$
where $W^{-1}$ is a regularized inverse. Since we know the exact dipole density we can choose the regularization parameter in an optimal way. Using 
truncated singular-value regularization (TSVD) with 110 singular values from total 186 we obtain very good results as shown in Figure B plotted in red, compared with the exact dipole density values plotted on the blue curve. The normalized RMS error is 0.01. In the case of real data from living hearts a good eye of the medical doctor is required to find out the optimal regularization parameter. If we use Tikhonov regularization instead of TSVD we find no statistically significant difference in the resultant calculated dipole density. If the basket is not placed in the center the results get worse, but not dramatically. However using only 48 electrodes instead of 186 gives poor results showing that we have a large discretization error in this case. The remedy in view of the real situation with the AcQMap system is interpolation of the measured potential values.
For this interpolation on a triangular surface the method of Oostendorp, van Oosterom and Huiskamp [10] is very useful, because it minimizes $\Delta V$. This is the best strategy because the exact potential would satisfy Laplace's equation $\Delta V=0$. .

\section{Phase dynamics}

Cardiac fibrillation is the main cause of death in the western world. Nevertheless its underlying mechanisms of activation are still poorly understood. Obviously mapping of cardiac potentials is not sensitive enough to improve the situation. There is considerable hope that dipole density maps can help. These maps show the amplitude of the dipole density $d(t,x)$ distributed over the heart wall ($x\in S$) as a function of time $t$. But in addition to the amplitude the phase of the dipole density gives important information as it is the case in the phase analysis of electrograms [11-14] (and references given there).

To define the phase, the dipole density $d$ is considered as the real part of a complex function whose imaginary part is given by the Hilbert transform
$$(Hd)(t)={1\over\pi}P\int\limits_{-\infty}^{+\infty}{d(t')\over t-t'}dt'\eqno(4.1)$$
where $P$ stands for the principle value integral.
The phase $\Phi(t)$ is then equal to the phase of the complex number $d+iHd$, that means
$$\Phi(t)=\arctan{-d\over Hd}.\eqno(4.2)$$
If the phase moves out of the interval $[-\pi/2, \pi/2]$ it must be continued continuously until the full period $[-\pi, \pi]$ is reached. This arctan-function is denoted by arctan2 in Matlab so that the general definition is
$$\Phi(t)=\arctan 2(-d, Hd).\eqno(4.3)$$
In this definition we have assumed that the mean value of $d(t)$ over time is zero. By adding or subtracting $2\pi$, $\Phi(t)$ can be made continuous in $t$. 

Since the phase can be calculated at every point $x$ where the dipole density $d(t,x)$ has been determined, we actually get a phase map $\Phi(t,x)$ on the heart wall for every instant $t$. This map shows singular points where the phase is undetermined. Such a phase singularity is actually a singularity of the gradient of $\Phi(t,x)$. In fact, iif we integrate $\nabla_x\Phi(t,x)$ along a closed curve we get zero, except some singularity of the gradient is included. This is the same situation as in 2-dimensional hydrodynamics where the flow velocity integrated along a closed curve give the circulation which vanishes except a vertex is included. Considering $\vec v(t,x)=\nabla_x\Phi(t,x)$ as a flow velocity we get a complete hydrodynamical analogy. We have a potential flow, the phase is the velocity potential. In hydrodynamics the circulation is conserved in the course of time. We want to investigate the same property for the phase singularities on the heart wall.

The first observation is that the phase singularities are quantized vortices. That means the contour integral of $\nabla_x\Phi(t,x)$ (the circulation) always has the same value $\pm 2\pi$. Studying various phase maps on the heart wall we have found that the vortices always appear in pairs: one with circulation $+2\pi$ and a second one with circulation $-2\pi$. This shows that the circulation is indeed conserved like in hydrodynamics.
One vortex cannot appear or disappear suddenly, it can only annihilate together with a partner of opposite circulation. In the healthy heart there seems to exist one pair of rather stable vortices only. The two vortices can be joined by a line where the flow velocity $\vec v(x)$ is maximal (see Figure 2). This line may be considered as the activation front. The front is most easily found by a jump from $+\pi$ to $-\pi$ in the phase.  During one heart beat this activation front moves over the heart wall, while its endpoints at the vortices remain more or less fixed. If some arrhytmia is developed, more and more vortex pairs appear and move around. Beside vortices sometimes sources and perhaps also sinks show up which have a closed activation front. It is clear that the study of this flow dynamics will be an important tool for understanding complicated arrhytmias.

\begin{figure}
\centering
\includegraphics[width=\textwidth]{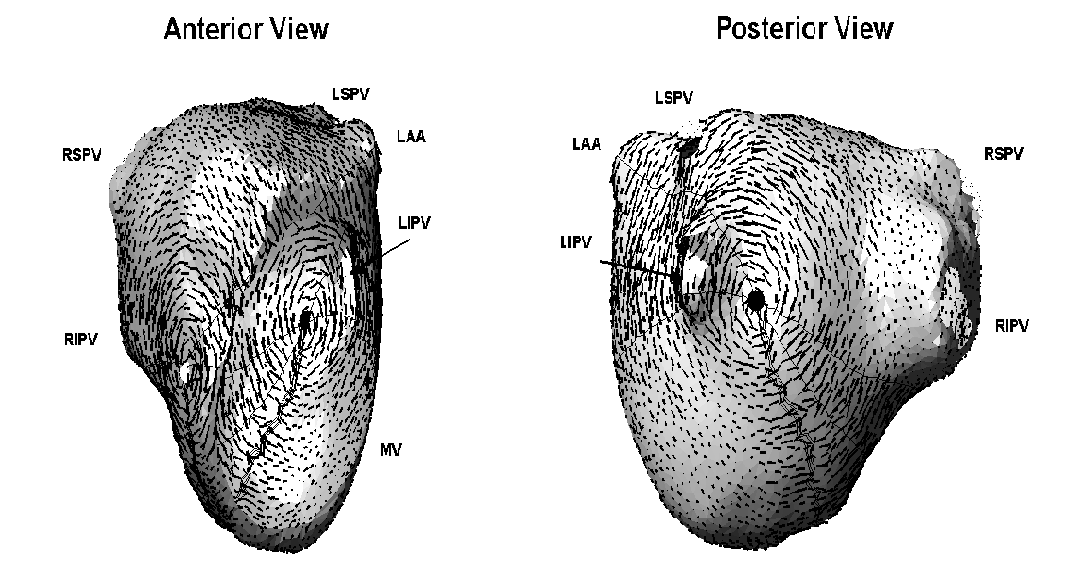}
\caption{Human left atrium with a pair of $\pm$ vertices (black and white) together with the phase flow. The black line is the activation front. LAA: Left Atrial Appendage, LSPV: Left Superior Pulmonary Vein, LIPV: Left Inferior Pulmonary Vein, RSPV: Right Superior Pulmonary Vein, RIPV: Right Inferior Pulmonry Vein, MV: Mitral Valve. }
\end{figure}

{\bf Acknowledgment}

We thank Graydon Beatty from Acutus medical for innumerable elucidating discussions and communication of information. Thanks are also due to other members of the Acutus team, in particular Min Zhu and Xinwei Shi and, of course, Randy Werneth.

\end{document}